\documentclass[12pt]{article}
\usepackage{amssymb}
\usepackage{amsmath,amsfonts,feynmp}
\usepackage{tipa}
\usepackage{stmaryrd}

\makeatletter\@addtoreset{equation}{section}\makeatother

\def\be{\begin{equation}}
\def\ee{\end{equation}}
\def\bea{\begin{eqnarray}}
\def\eea{\end{eqnarray}}

\makeatletter\@addtoreset{equation}{section}\makeatother

\renewcommand{\title}[1]{\vbox{\center\LARGE{#1}}\vspace{5mm}}
\renewcommand{\author}[1]{\vbox{\center#1}\vspace{5mm}}
\newcommand{\address}[1]{\vbox{\center\em#1}}

\begin{document}

\unitlength = .8mm

\begin{titlepage}
\begin{center}
\hfill \\
\hfill \\
\vskip 1.5cm

\title{On the $z=4$ Ho${\rm\check{r}}$ava-Lifshitz Gravity }

\vskip 0.5cm
 {Rong-Gen Cai\footnote{Email: cairg@itp.ac.cn}, Yan Liu\footnote{Email: liuyan@itp.ac.cn}} and {Ya-Wen
Sun\footnote{Email: sunyw@itp.ac.cn}}

\address{ Key Laboratory of Frontiers in Theoretical Physics
\\ Institute of Theoretical Physics, Chinese Academy of Sciences,
\\P.O. Box 2735, Beijing 100190, China}

\end{center}

\vskip 1cm

\abstract{ We consider $z=4$ Ho${\rm\check{r}}$ava-Lifshitz gravity
in both 3+1 and 4+1 dimensions. We find black hole solutions in the
IR region for a kind of $z=4$ Ho${\rm\check{r}}$ava-Lifshitz gravity
which is inherited from the new massive gravity in three dimensions
and an analog of the new massive gravity in four dimensions through
the quantum inheritance principle. We analyze thermodynamic
properties for the black hole solutions for $z=4$
Ho${\rm\check{r}}$ava-Lifshitz gravity. We also write out the
Friedmann equation in 3+1 dimensions for cosmological solutions.}

\vfill

\end{titlepage}


\section{Introduction}
Recently, Ho${\rm\check{r}}$ava-Lifshitz gravity
\cite{{Horava:2008jf},{Horava:2008ih},Horava:2009uw} has attracted
much attention as a candidate quantum field theory of gravity with
$z=3$ in the UV, where $z$ measures the degree of anisotropy between
space and time. In 3+1 dimensions, the
Ho${\rm\check{r}}$ava-Lifshitz theory has a $z=3$ fixed point in the
UV and flows to a $z=1$ fixed point in the IR, which is just the
classical Einstein-Hilbert gravity theory. This theory is
renormalizable in the sense that the effective coupling constant is
dimensionless in the UV.

Much work has been done in Ho${\rm\check{r}}$ava-Lifshitz gravity
theory \cite{{Horava:2009if},{Takahashi:2009wc},{Calcagni:2009ar},
{Kiritsis:2009sh},{Kluson:2009sm},{Lu:2009em},{Mukohyama:2009gg},{Brandenberger:2009yt},{Nikolic:2009jg},{Nastase:2009nk},{Cai:2009pe},{Piao:2009ax},{Gao:2009bx}}.
In
\cite{{Takahashi:2009wc},{Calcagni:2009ar},{Kiritsis:2009sh},{Mukohyama:2009gg},{Brandenberger:2009yt},{Piao:2009ax},{Gao:2009bx}},
 the possible effect of this quantum gravity from the
sky was studied. In \cite{Lu:2009em,{Nastase:2009nk},{Cai:2009pe}}
the solutions for $z=3$ Ho${\rm\check{r}}$ava-Lifshitz gravity in
3+1 dimensions have been considered and in \cite{Cai:2009pe} the
thermodynamics of black hole solutions was discussed. In
\cite{Nikolic:2009jg} the time problem in this quantum gravity
theory was considered. The focus of all these work was on the
Ho${\rm\check{r}}$ava-Lifshitz theory with $z=3$ in the UV, while
Ho${\rm\check{r}}$ava-Lifshitz theory with $z=4$ in the UV can also
be constructed from the curvature square terms in the three spatial
dimensions \cite{Horava:2009uw}, which is power-counting
super-renormalizable in 3+1 dimensions.

There are several reasons that the $z=4$
Ho${\rm\check{r}}$ava-Lifshitz theory is of much importance. First,
in \cite{Horava:2009if} it was shown that the spectral dimension
calculated using the numerical Causal Dynamical Triangulations (CDT)
approach \cite{Ambjorn:2005db} for quantum gravity in 3+1 dimensions
prefers the $z=4$ Ho${\rm\check{r}}$ava-Lifshitz gravity in the UV.
Second, in \cite{Horava:2009uw} it was pointed out that in the case
when the scalar mode of the metric is present as a physical field,
in order for the theory to be power-counting renormalizable for
general value of $\lambda$, which is a dimensionless coupling
defined below, one way is to add super-renormalizable terms. Third,
from the conjecture of the quantum inheritance principle, the
construction of a four dimensional renormalizable gravity is
intimately related to a three dimensional renormalizable
relativistic theory. The potential of the $z=3$
Ho${\rm\check{r}}$ava-Lifshitz theory in the UV can be constructed
from the topological massive theory (TMG)
\cite{{Deser:1981wh},{Deser:1982vy},Li:2008dq} in three dimensions
by the so called ``detailed balance condition.'' There is also
another renormalizable relativistic massive gravity in three
dimensions, which is proposed recently by \cite{Bergshoeff:2009hq}
and further investigated in
\cite{{Nakasone:2009bn},{Clement:2009gq},{Liu:2009bk},{Nakasone:2009vt},{Liu:2009kc},{Liu:2009ph},{AyonBeato:2009yq},Oda:2009ys}.
This new massive gravity (NMG) can be used to construct a $z=4$
Ho${\rm\check{r}}$ava-Lifshitz theory. Thus in the rest of this
paper we will focus our attention on $z=4$
Ho${\rm\check{r}}$ava-Lifshitz gravity .

In this paper, we would like to consider $z=4$ Ho${\rm\check{r}}$ava
gravity in both 3+1 and 4+1 dimensions. We will first write out the
form of the action for the $z=4$ Ho${\rm\check{r}}$ava-Lifshitz
gravity in 3+1 dimensions constructed from general $R^2$ terms in
three dimensions. Then we will search for the topological black hole
solutions of the theory inherited from the new massive gravity and
discuss thermodynamic properties of the black hole solutions. We
also get the Friedmann equation in 3+1 dimensions for cosmological
solutions. In 4+1 dimensions, $z=4$ Ho${\rm\check{r}}$ava-Lifshitz
gravity is renormalizable by power counting and the general form of
the action is similar to the case of 3+1 dimensions. We also search
for black hole solutions of the theory constructed from an analog of
the new massive gravity in four dimensions and analyze the
thermodynamic properties of these black hole solutions.

Our note is organized as follows. In Sec.~2 we will formulate the
$z=4$ Ho${\rm\check{r}}$ava gravity in 3+1 dimensions and discuss
its black hole solutions as well as cosmological solutions. In
Sec.~3 we will formulate the $z=4$ Ho${\rm\check{r}}$ava-Lifshitz
gravity in 4+1 dimensions and study its solutions. Sec.~4 is devoted
to conclusions and discussions.

\section{$z=4$ Ho${\rm\check{r}}$ava-Lifshitz gravity in 3+1 dimensions}
In this section, we will give the form of the action for $z=4$
Ho${\rm\check{r}}$ava-Lifshitz gravity in 3+1 dimensions and search
for solutions for the Ho${\rm\check{r}}$ava-Lifshitz gravity
inherited from the new massive gravity.

\subsection{Brief review of Ho${\rm\check{r}}$ava-Lifshitz gravity}

First we will review the construction of
Ho${\rm\check{r}}$ava-Lifshitz gravity in $D+1$
dimensions~\cite{Horava:2008ih,{Horava:2009uw}}. Similar to the ADM
decomposition of the metric in general relativity, we can write the
$D+1$ dimensional metric as \be
ds^2=-N^2c^2dt^2+g_{ij}(dx^i-N^idt)(dx^j-N^jdt), \ee where
$i=1,\cdots,D$ and $c$ is the speed of light. The scaling dimension
is modified at the fixed point with Lifshitz index $z$ as
$t\rightarrow b^z t, x^i\rightarrow b x^i,$ under which $g_{ij}$ and
$N$ are invariant while $N^i\rightarrow b^{1-z}N^i$. In the
following we measure the scaling properties in the units of inverse
spatial length, so we have $[t]=-z, ~[x^i]=-1, ~[c]=[N^i]=z-1$ at
the fixed point. When $z=1$, the scaling properties get back to our
familiar relativistic case, and this is the IR limit. In the UV
region, the Ho${\rm\check{r}}$ava-Lifshitz theory switches to other
$z$ to make this theory renormalizable.

The kinetic term is given by \be S_K=\frac{2}{\kappa^2}\int dtd^Dx
\sqrt{g}N(K_{ij}K^{ij}-\lambda K^2),\ee where
$K_{ij}=\frac{1}{2N}(\dot{g}_{ij}-\nabla_iN_j-\nabla_jN_i)$ and the
scaling dimension of the coupling constant $\kappa$ at the fixed
point is $[\kappa]=\frac{z-D}{2}$. Here $\lambda$ is a dimensionless
parameter and it is equal to 1 in the IR to restore general
relativity. In the UV, to make sure that our theory is
power-counting renormalizable we should have $z_{{\rm UV}}\geq D$.
When $z_{{\rm UV}}=D$, the theory is renormalizable and when
$z_{{\rm UV}}>D$, the theory is super-renormalizable.

The potential term is obtained by the ``detailed-balance principle"
of the form \be S_V=\frac{\kappa^2}{8}\int dt d^Dx \sqrt{g}N
E^{ij}{\mathcal{G}}_{ijk\ell} E^{k\ell}\ee with $E^{ij}$ coming from
a $D$ dimensional relativistic action in the form \be\label{dbc}
E^{ij}=\frac{1}{\sqrt{g}}\frac{\delta W_D[g_{ij}]}{\delta
g_{ij}},\ee and \be{\mathcal
{G}}_{ijk\ell}=\frac{1}{2}(g_{ik}g_{j\ell}+g_{i\ell}g_{jk})-\tilde{\lambda}g_{ik}g_{j\ell},
~~\tilde{\lambda}=\frac{\lambda}{D\lambda-1}. \ee Because we
consider the spatial isotropic theory, $W_D$ must be the Euclidean
action of a relativistic theory. The expression (\ref{dbc}) connects
a $D$-dimensional system described by the action $W_D$ to a
$D+1$-dimensional system described by the action $S_K-S_V.$ This is
the so called ``detailed balance condition." It's well-known that it
plays an important role in the quantum field theory for scalar
fields and the role in the quantum field theory for gauge fields has
also been studied in~\cite{Horava:2008jf}.

\subsection{Action for $z=4$ Ho${\rm\check{r}}$ava-Lifshitz gravity in 3+1
dimensions}

To get the form of the action of the $z=4$
Ho${\rm\check{r}}$ava-Lifshitz gravity in 3+1 dimensions from the
detailed balance condition we follow \cite{Horava:2009uw} to begin
from the action for general higher derivative corrections in three
dimensional gravity theory
 \bea W[g_{ij}]&=&W_1+W_2+W_3,\\
W_1&=&\mu\int d^3x\sqrt{g}(R-2\Lambda_{W}),\\
W_2&=&\frac{1}{w^2}\int d^3x\sqrt{g}\varepsilon^{ijk}\Gamma^m_{il}
[\partial_j\Gamma^l_{km}+\frac{2}{3}\Gamma^l_{jn}\Gamma^n_{km}],\\
W_3&=&\frac{1}{M}\int d^3x\sqrt{g}(R_{ij}R^{ij}+\beta R^2). \eea
Here $\mu$, $\omega^2$, $M$ and $\beta$ are coupling constants. We
take $\varepsilon_{ijk}=\sqrt{g}\epsilon_{ijk}$ with
 $\epsilon_{123}=1$. When we only have $W_1$ and $W_2$, $W$ is the action for TMG and the resulting
Ho${\rm\check{r}}$ava-Lifshitz gravity just has $z=3$ in the UV. To
get $z=4$ we need to keep the $W_1$ and $W_3$ terms and in this case
when we choose $\beta=-3/8$, $W$ represents the action of the
Euclidean version of the NMG theory. Because the NMG itself is
renormalizable in the Minkowski spacetime~\cite{Oda:2009ys}, we
believe that the 3+1 dimensional $z=4$ theory constructed at this
special $\beta$ should be much simpler than at other $\beta$, which
is indeed the case when we search for solutions in the next
subsection. When we have all the $W_1$, $W_2$ and $W_3$ terms at
$\beta=-3/8$, the action $W$ reduces to the Euclidean version of the
generalized massive gravity theory (GMG) which is the generalized
version of three-dimensional massive gravity theory
\cite{{Bergshoeff:2009hq},Liu:2009ph}.

Now we follow the quantum inheritance principle and the ``detailed
balance principle" to formulate the action of the 3+1 dimensional
theory. We will keep $\beta$ unfixed in this procedure. We can get
$E^{ij}$ from (\ref{dbc}) as
 \bea E^{ij}&=&\frac{1}{\sqrt{g}}\frac{\delta W}{\delta
{g_{ij}}}=\frac{2}{\omega^2}C^{ij}-\mu(G^{ij}+\Lambda_W
g^{ij})-\frac{1}{M}L^{ij},\nonumber\\
C^{ij}&=&\varepsilon^{ik\ell}\nabla_k(R^j_{~\ell}-\frac{1}{4}R\delta^j_{~\ell}),\nonumber\\
G^{ij}&=&R^{ij}-\frac{1}{2}g^{ij}R,\nonumber\\
L^{ij}&=&(1+2\beta)(g^{ij}\nabla^2-\nabla^i\nabla^j)R+\nabla^2G^{ij}\nonumber\\&&~~+2\beta
R(R^{ij}-\frac{1}{4}g^{ij}R)+
2(R^{imjn}-\frac{1}{4}g^{ij}R^{mn})R_{mn}.\eea By combining the
kinetic terms and the potential terms, we obtain the action
\bea\label{action}{\mathcal {L}}&=&{\mathcal {L}}_0+{\mathcal
{L}}_1,\nonumber\\
{\mathcal{L}}_0&=&\sqrt{g}N\big\{\frac{2}{\kappa^2}(K^{ij}K_{ij}-\lambda
K^2)+
\frac{\kappa^2\mu^2(\Lambda_WR-3\Lambda_W^2)}{8(1-3\lambda)}\big\},\nonumber\\
{\mathcal{L}}_1&=&-\sqrt{g}N\frac{\kappa^2}{8}\Big\{\frac{4}{\omega^4}C^{ij}C_{ij}-\frac{4\mu}{\omega^2}C^{ij}R_{ij}
-\frac{4}{\omega^2M}C^{ij}L_{ij}+\mu^2G_{ij}G^{ij}+\frac{2\mu}{M}G^{ij}L_{ij}\nonumber\\&&~+\frac{2\mu}{M}\Lambda_W
L+\frac{1}{M^2}L^{ij}L_{ij}-\tilde{\lambda}\big(\frac{L^2}{M^2}-\frac{\mu
L}{M}(R-6\Lambda_W)+\frac{\mu^2}{4}R^2\big)\Big\},\eea
 where \be
L\equiv
g^{ij}L_{ij}=\bigg(\frac{3}{2}+4\beta\bigg)\nabla^2R+\frac{\beta}{2}R^2+\frac{1}{2}R_{ij}R^{ij}.\ee
When $\beta=-3/8$ and $w\rightarrow\infty$, the action
(\ref{action}) just represents the $z=4$
Ho${\rm\check{r}}$ava-Lifshitz gravity inherited from the new
massive gravity, and this is the case we are most interested in
because the renormalizability of the new massive gravity in
Minkowski spacetime has been confirmed in \cite{Oda:2009ys}.

In the UV, the terms with the highest scaling dimension in the
potential terms will dominate, i.e. the terms inherited from $W_3$,
and the theory exhibits a $z=4$ Lifshitz type fixed point in the UV
region. In the IR, the terms with the lowest scaling dimension in
the potential term will dominate, i.e. ${\mathcal{L}}_0$ will
dominate. In order to obtain general relativity in the IR region,
the effective coupling should be related to the speed of light $c$,
 the Newton coupling $G$ and the effective cosmological constant as \be
c=\frac{\mu\kappa^2}{4}\sqrt{\frac{\Lambda_W}{1-3\lambda}},\ee \be
G_N=\frac{\kappa^2c}{32\pi }\ee and \be
\Lambda=\frac{3}{2}\Lambda_W.\ee From the expression of the speed of
light we see that in the IR region the cosmological constant can
only be negative.
\subsection{Black hole solutions}

To search for black hole solutions in this theory, we follow
\cite{Lu:2009em,{Cai:2009pe}} to assume the form of the metric as
\be\label{metric1} ds^2=-
\tilde{N}^2(r)f(r)c^2dt^2+\frac{dr^2}{f(r)}+r^2d\Omega^2_k,\ee where
$d\Omega^2_k$ denotes the line element for a two dimensional
Einstein space with constant scalar curvature $2k$ and volume
$\Omega_k$. Without loss of generality, we take $k=0,~\pm 1$
respectively. Then we can substitute this into the action
(\ref{action}) to get the equation of motion for $\tilde{N}(r)$ and
$f(r)$ at any given energy scale. It is very complicated to find
explicit solutions for a general $\beta$ and we will focus on our
most interesting case: $\beta=-3/8$, which is the
Ho${\rm\check{r}}$ava-Lifshitz theory related to NMG in three
dimensions. Coincidentally we find that at $\beta=-3/8$, the action
has the least number of contributions which makes the calculation
simplified. In this case when the Cotton tensor terms are also
included, the three dimensional theory is the Euclidean generalized
massive gravity theory (GMG), which has been recently studied in
\cite{{Bergshoeff:2009hq},{Nakasone:2009bn},{Clement:2009gq},{Liu:2009bk},{Nakasone:2009vt},{Liu:2009kc},{Liu:2009ph},{AyonBeato:2009yq},Oda:2009ys}.
In addition, this GMG theory is renormalizable. As we know that the
norm of the ground state of a $D+1$ dimensional Lifshitz theory is
related to the partition function of the $D$ dimensional
relativistic theory \cite{Horava:2008ih}, it is natural to consider
this theory seriously.

We choose $\lambda=1$ to seek for solutions in the IR region. Then
we substitute the metric ansatz (\ref{metric1}) into the action
(\ref{action}) and find that \bea
I&=&\frac{{\kappa}^2\Omega_k}{16\sqrt{-\Lambda_W^3}}\int dtdx
\tilde{N}\Big(\tilde{\mu}^2[x^3-2x(f-k)+\frac{(f-k)^2}{x}]\nonumber\\&&~~-2\tilde{\beta}\tilde{\mu}[\frac{(f-k)^3}{x^3}
-\frac{(f-k)^2}{x}]+\tilde{\beta}^2\frac{(f-k)^4}{x^5}\Big)^{'},\eea
where we have defined $\tilde{\mu}=-\mu \Lambda_W,
\tilde{\beta}=\frac{\Lambda_W^2}{4M}$ and $x=\sqrt{-\Lambda_W}r$,
and the prime denotes the derivative with respect to $x$. From the
action, we can obtain the equations of motion as \bea
0&=&\tilde{N}^{'}\big(\tilde{\beta}\frac{(f-k)^2}{x^2}-\tilde{\mu}(f-k-x^2)\big)
\big(\frac{2\tilde{\beta}(f-k)}{x^2}-\tilde{\mu}\big),\nonumber\\
c_0&=&\frac{1}{x}\Big(\tilde{\beta}\frac{(f-k)^2}{x^2}-\tilde{\mu}(f-k-x^2)\Big)^2,\eea
where $c_0\geq0$ is an integration constant. When $c_0>0$, from the
equations of motion we can see that $N(r)$ should be a constant,
which can always be set to $1$ by a redefinition of the coordinate
$t$. The function $f(r)$ can be obtained as \be \label{solution}
f=k+\frac{\tilde{\mu}}{2\tilde{\beta}}x^2\bigg(1
\pm\sqrt{1-\frac{4\tilde{\beta}}{x^2\tilde{\mu}^2}(\tilde{\mu}
x^2\pm\sqrt{c_0x})}\bigg).
 \ee
 To reduce the theory to the $z=3$ one when
$\tilde{\beta}$ goes to zero, the two $\pm$ in the solution
(\ref{solution}) can only be chosen to be $-$. Thus the solution
becomes \be\label{solution2}
f=k+\frac{\tilde{\mu}}{2\tilde{\beta}}x^2\bigg(1
-\sqrt{1-\frac{4\tilde{\beta}}{x^2\tilde{\mu}^2}(\tilde{\mu}
x^2-\sqrt{c_0x})}\bigg).\ee In that case, we have to impose
$4\tilde{\beta}/\tilde{\mu}\leq1$ in order to have a well-defined
vacuum solution.  This solution is asymptotically $AdS_4$ and when
$\tilde{\beta}$ goes to zero, this just comes back to the solution
given in \cite{Cai:2009pe} up to a redefinition of the integration
constant. When $c_0$ is zero, $\tilde{N}(r)$ can be an arbitrary
function of $r$ and $f(r)$ is just (\ref{solution}) with $c_0=0$. We
suspect that other branches in (\ref{solution}) are unstable
perturbatively. To confirm this, however, a careful analysis is
needed. There is a potential singularity at the point where the
square root vanishes in (\ref{solution2}), in addition to the one at
$x=0$.

\subsection{Thermodynamic properties of the black holes}

The black hole horizon $x_+$ is given by the largest root of the
equation $f(r)=0$. The mass of the solution can be obtained by the
Hamiltonian approach following \cite{CS,BTZ}
\begin{equation}
m= \frac{\kappa^2 \Omega_k}{16(-\Lambda_W)^{3/2}} c_0,
\end{equation}
while $c_0$ can be expressed in terms of the horizon radius
\begin{equation}
 c_0 =\frac{1}{x_+}\left (\frac{{\tilde \beta} k^2}{x_+^2} +\tilde \mu
 (k+x_+^2)\right )^2.
 \end{equation}
 The Hawking temperature associated with the black hole is found
 to be
 \begin{equation}
 T = \frac{\sqrt{-\Lambda_W}}{8\pi x_+} \frac{3x_+^2-k-
 \frac{5\tilde \beta k^2}{\tilde \mu x_+^2}}{1+\frac{2\tilde \beta
 k}{\tilde \mu x_+^2}}.
 \end{equation}
 When $\tilde \beta \to 0$, it goes back to the temperature of
 black hole in the $z=3$ Ho${\rm\check{r}}$ava-Lifshitz
 gravity~\cite{Cai:2009pe}. Following \cite{Cai:2009pe}, one can
 obtain the black hole entropy by integrating the first law of
 black hole thermodynamics, which gives
 \begin{equation}
S = \frac{\pi\kappa^2\mu^2\Omega_k}{4} \left( x_+^2 +2k \ln x_+
-\frac{3\tilde \beta k^2}{\tilde \mu x_+^2} -\frac{\tilde \beta^2
k^3}{\tilde \mu^2 x_+^4}  +\frac{4\tilde \beta k}{\tilde \mu}\ln
x_+\right) +S_0.
 \end{equation}
Here $S_0$ is an integration constant, which cannot be determined by
the thermodynamic method. Note that when $\tilde \beta \to 0$, the
entropy reduces to the one given in \cite{Cai:2009pe}. The leading
term is the area term, and the second is a logarithmic correction.
Clearly one can see from the black hole entropy that $z=4$ terms in
the gravity action lead to the entropy correction terms $1/A$ and
$1/A^2$, besides a new logarithmic term, here $A\sim x_+^2$ is the
horizon area of the black hole. Note that for a Ricci flat black
hole with $k=0$, only the area term remains in the entropy
expression.

\subsection{Cosmological solutions}
In this subsection, we are going to write out the Friedmann equation
for cosmological solutions in the $z=4$
Ho${\rm\check{r}}$ava-Lifshitz gravity theory in $3+1$ dimensions.
We first assume that the Friedmann-Robertson-Walker (FRW) metric is
of the form \be\label{frw}
ds^2=-N^2(t)c^2dt^2+a^2(t)\Big(\frac{dr^2}{1-kr^2}+r^2(d\theta^2+{\rm
sin}^2\theta d\phi^2)\Big),\ee where $k=1,0,-1$ corresponding to a
closed, flat and open universe respectively.

We substitute the ansatz ($\ref{frw}$) into the action
(\ref{action}), and obtain \be I_g=\int dtd^3x
\sqrt{g}\Big\{-\frac{6(-1+3\lambda)\dot{a}^2}{\kappa^2Na^2}+\frac{3\kappa^2N}{8(-1+3\lambda)a^8}
\big(2(1+3\beta)\frac{k^2}{M}-k\mu a^2+\Lambda \mu
a^4\big)^2\Big\},\ee where the dot denotes the derivative to $t$.
Supposing that the matter contribution is equivalent to an ideal
fluid and satisfies the ``separated-detailed balance
condition"~\cite{Calcagni:2009ar}, we have $\rho$ and $p$ of the
matter to be \be \rho=-\frac{1}{\sqrt{g}}\frac{\delta
I_{\phi}}{\delta {N}},~~~ p=\frac{a}{3N\sqrt{g}}\frac{\delta
I_{\phi}}{\delta {a}}.\ee The equation of motion for $N(t)$ gives
\be\label{nt}
\frac{6}{\kappa^2}(-1+3\lambda)H^2+\frac{3\kappa^2}{8(-1+3\lambda)a^8}\big(
2(1+3\beta)\frac{k^2}{M}-k\mu a^2+\Lambda_W \mu a^4\big)^2=\rho,\ee
and the equation of motion for $a(t)$ gives \bea\label{at}
\frac{4}{\kappa^2}(-1&+&3\lambda)(\dot{H}+\frac{3}{2}H^2)+\frac{\kappa^2}{4(-1+3\lambda)a^8}\big(
2(1+3\beta)\frac{k^2}{M}-k\mu a^2\nonumber\\&&+\Lambda_W \mu
a^4\big)\big(-5(1+3\beta)\frac{k^2}{M}+\frac{1}{2}k\mu
a^2+\frac{3}{2}\Lambda_W \mu a^4\big)=-p,\eea where \be
H\equiv\frac{\dot{a}}{Na},~~~\dot{
H}\equiv\frac{1}{N}\partial_t\Big(\frac{\dot{a}}{Na}\Big). \ee Form
(\ref{nt}) and (\ref{at}), we can obtain \be
\dot{\rho}+3H(\rho+p)=0.\ee

Note that for $k=0$, there is no contribution from the higher-order
derivative terms of the action. For $k=\pm1$, the contribution of
higher derivative terms dominates for small $a$ and can be ignored
at large $a$. This behavior is very analogous to the $z=3$ case
\cite{{Calcagni:2009ar}, {Kiritsis:2009sh},
Lu:2009em,{Brandenberger:2009yt}} and indeed, when $M\rightarrow
\infty$, we come back to the $z=3$ case
\cite{Kiritsis:2009sh,{Lu:2009em}}. We choose the gauge $N(t)=1$ and
for vacuum solutions with $\rho=p=0$, in order to obtain de-Sitter
solutions without matter, we should perform an analytic continuation
of the parameters~\cite{Lu:2009em}, \be \mu\rightarrow
i\mu,~~~\omega^2\rightarrow -i\omega^2,~~~ M\rightarrow -iM,\ee then
$c=\frac{\mu\kappa^2}{4}\sqrt{\frac{\Lambda_W}{3\lambda-1}}$ and the
Friedmann equations are \be
\Big(\frac{\dot{a}}{a}\Big)^2=\frac{\kappa^4}{16(-1+3\lambda)^2}\big(2(1+3\beta)\frac{k^2}{Ma^4}-\frac{k\mu}
{a^2}+\Lambda_W \mu \big)^2.\ee This equation can be integrated to
give the explicit solution of $a(t)$ and because the solution is
long and complicated, we don't write it out here. When $M\rightarrow
\infty$, it comes back to the $z=3$ case and it can be easily shown
that for $k=0$, it has de Sitter solutions.

\section{$z=4$ Ho${\rm\check{r}}$ava-Lifshitz gravity in 4+1 dimensions}

In this section, we will study the $z=4$
Ho${\rm\check{r}}$ava-Lifshitz gravity in 4+1 dimensions. The $z=4$
theory in 4+1 dimensions is power-counting renormalizable. If the
Ho${\rm\check{r}}$ava-Lifshitz gravity can indeed be a quantum
gravity, it is interesting to investigate the $AdS_5$/$CFT_4$
correspondence in the framework of this theory. Thus it is
interesting to study the $z=4$ Ho${\rm\check{r}}$ava-Lifshitz
gravity in 4+1 dimensions.

\subsection{Action}
The general four dimensional relativistic Lagrangian with higher
derivative terms is of the form \be\label{w} W=\mu\int
d^4x\sqrt{g}(R-2\Lambda_W)+\frac{1}{M}\int
d^4x\sqrt{g}(R_{ij}R^{ij}+\beta R^2).\ee Here the second term is the
most general curvature square contribution because the Gauss-Bonnet
term is a topological invariant in four dimensions.

Using (\ref{w}), the $E^{ij}$ can be obtained \bea
E^{ij}&=&-\mu(G^{ij}+\Lambda_W
g^{ij})-\frac{1}{M}L^{ij},\nonumber\\
G^{ij}&=&R^{ij}-\frac{1}{2}g^{ij}R,\nonumber\\
L^{ij}&=&(1+2\beta)(g^{ij}\nabla^2-\nabla^i\nabla^j)R+\nabla^2G^{ij}\nonumber\\&&~~+2\beta
R(R^{ij}-\frac{1}{4}g^{ij}R)+
2(R^{imjn}-\frac{1}{4}g^{ij}R^{mn})R_{mn}.\eea Then the Lagrangian
of the $z=4$ Ho${\rm\check{r}}$ava-Lifshitz gravity in 4+1
dimensions are of the following form \bea{\mathcal {L}}&=&{\mathcal
{L}}_0+{\mathcal
{L}}_1,\nonumber\\
{\mathcal{L}}_0&=&\sqrt{g}N\{\frac{2}{\kappa^2}(K^{ij}K_{ij}-\lambda
K^2)+
\frac{\kappa^2\mu^2(\Lambda_WR-2\Lambda_W^2)}{4(1-4\lambda)}\},\nonumber\\
{\mathcal{L}}_1&=&-\sqrt{g}N\frac{\kappa^2}{8}\Big\{\mu^2G_{ij}G^{ij}
+\frac{2\mu}{M}G^{ij}L_{ij}+\frac{2\mu}{M}\Lambda_W
L+\frac{1}{M^2}L^{ij}L_{ij}\nonumber\\&&~-\tilde{\lambda}(\frac{L^2}{M^2}-\frac{2\mu
L}{M}(R-4\Lambda_W)+\mu^2R^2)\Big\},\eea where \be
L=2(1+3\beta)\nabla^2R.\ee In the UV, this theory will exhibit a
$z=4$ Lifshitz-type fixed point and in the IR, the ${\mathcal{L}}_0$
term dominates and the theory flows to the $z=1$ one. In order to
get back to general relativity in the IR region, the effective
couplings should be related to the speed of light $c$, the Newton
coupling $G$ and the effective cosmological constant $\Lambda$ at IR
in the way \be
c=\frac{\mu\kappa^2}{\sqrt{8}}\sqrt{\frac{\Lambda_W}{1-4\lambda}},\ee
\be G_N=\frac{\kappa^2c}{32\pi }\ee and  \be \Lambda=\Lambda_W,\ee
respectively. In the IR region, to get back to general relativity
$\lambda$ should be chosen to be $1$ and $\Lambda_W$ can only be
negative to have a physical speed of light.

\subsection{Solutions}
In this subsection we will seek for black hole solutions in the IR
region. We make an ansatz for the  metric form as \be\label{metric2}
ds^2=-
\tilde{N}^2(r)f(r)c^2dt^2+\frac{dr^2}{f(r)}+r^2d\Omega^2_k,\ee where
$d\Omega^2_k$ is the three-dimensional Einstein manifold with
constant scalar curvature $6k$, which we may choose to be $k=0,~\pm
1$, without loss of generality.

It is quite difficult to find the exact solution for a general
$\beta$ in this case, and
 motivated by the 3+1 dimensional case, we consider a special value
 of $\beta=-1/3$, which comes from an unsuccessful attempt to
 generalize the new massive gravity to the case in four dimensions
 \cite{Nakasone:2009bn}. We
solve the solution following the method of
\cite{{Lu:2009em},Cai:2009pe}. We substitute the metric ansatz
into the action and find that
 \bea
I=\frac{{\kappa}^2\mu^2\Omega_k}{24}\int
dtdx&&\tilde{N}(x)\frac{1}{x}\Big\{12(1-\lambda)(f-k)^2+6x(f-k)(-2
x+ (1+2\lambda)f^{'})\nonumber\\&&~~+x^2(4x^2 -6
xf^{'}+3(1-\lambda)f^{'2})\Big\},\eea where $x=\sqrt{-\Lambda_W}r$.
Then in the IR region, we further consider the case of $\lambda=1$
and obtain
 \be
I=\frac{{\kappa}^2\mu^2\Omega_k}{24} \int dtdx
\tilde{N}\Big(\big(3(f-k)-x^2\big)^2\Big)^{'}.\ee  The equations of
motion are \bea
0&=&\tilde{N}^{'}(3(f-k)-x^2),\nonumber\\
c_0&=&\big(3(f-k)-x^2\big)^2,\eea where $c_0\geq0$ is an integration
constant. When $c_0>0$, $\tilde{N}(r)$ should be a constant and we
can always choose it to be $1$ through a redefinition of the
coordinate $t$. The solution of $f(r)$ can be easily obtained as\be
\label{f} f=k+\frac{x^2}{3}\pm \sqrt{c_0}, \ee where $c_0>0$. When
$c_0=0$, $\tilde{N}(r)$ can be an arbitrary function of $r$ and
$f(r)$ is just the solution (\ref{f}) with $c_0=0$. In the
calculation of the thermodynamic properties in the next subsection,
we only focus on the solutions with $\tilde{N}(r)=1$ and $c_0\geq0$,
which are of interest.  In addition let us  notice that the solution
(\ref{f}) is the same as the topological black hole solution in the
five-dimensional Chern-Simons gravity~\cite{CS}, or in the
five-dimensional Gauss-Bonnet gravity with a special Gauss-Bonnet
coefficient~\cite{Cai}.

\subsection{Thermodynamic properties of the black holes}

Note that when $c_0=0$ and $\tilde N=1$, the solution (\ref{f}) is a
five-dimensional AdS space.  When $c_0>0$, the black hole horizon
exists at $x_+^2 = 3(\mp \sqrt{c_0}-k)$ if $\mp \sqrt{c_0}-k \ge 0$.
Therefore when $k=1$ or $k=0$, the black hole horizon exists only in
the minus branch in (\ref{f}), while it requires $\sqrt{c_0} <1 $ in
the plus branch if $k=-1$. Of course, in the case of $k=-1$, the
black hole horizon always exists for the minus branch. According to
the Hamiltonian approach~\cite{CS,BTZ}, the mass of the solution is
 \begin{equation}
 m = \frac{\kappa^2\mu^2\Omega_k}{24}c_0.
 \end{equation}
 Comparing the solution (\ref{f}) with the one in the Gauss-Bonnet gravity~\cite{Cai},
 we expect that the solution with the plus branch
 is unstable perturbatively. Therefore in what follows, we focus
 on the minus branch only.  In that case, the black hole horizon is at
 $x_+^2 = 3(\sqrt{c_0}-k)$.  One can see immediately that when
 $k=1$, there is a mass gap $c_0=1$; when $c_0<1$, black hole
 horizon does not exist. While $k=-1$, there is a minimal
 horizon radius $x_{+{\rm min}}=\sqrt{3}$.

 The Hawking temperature of the black hole is easy to obtain.
 Either directly calculating surface gravity at black hole
 horizon, or requiring the absence of conical singularity at
 black hole horizon in the Euclidean sector of the black hole
 solution gives
 \begin{equation}
 T= \frac{\sqrt{-\Lambda_W}}{6\pi}x_+.
 \end{equation}
 Clearly the temperature is a monotonically increasing function of
 horizon radius $x_+$. Therefore the  black hole is always
 thermodynamically stable.  Using the first law of black hole
 thermodynamics, $dm=TdS$, we obtain the black hole entropy
 \begin{equation}
 S =\frac{\pi
 \kappa^2\mu^2\Omega_k}{27\sqrt{-\Lambda_W}}\left(x_+^3+9kx_+\right) +S_0.
 \end{equation}
 The leading term is proportional to the horizon area of the black
 hole, while the linear term vanishes for a Ricci flat black hole
 with $k=0$.

\section{Conclusion and Discussion}
In this note, we formulated the action of the $z=4$
Ho${\rm\check{r}}$ava-Lifshitz gravity in both 3+1 and 4+1
dimensions following \cite{Horava:2009uw}, and found static black
hole solutions  in the IR region. We also analyzed thermodynamic
properties of the black hole solutions. We only find explicit black
hole solutions for $\lambda=1$ and $\beta=-3/8$ in 3+1 dimensions
and black hole solutions for $\lambda=1$  and $\beta=-1/3$ in 4+1
dimensions. These special value of $\beta$ correspond to $z=4$
Ho${\rm\check{r}}$ava-Lifshitz gravity theories inherited from NMG
or GMG, which are of interest. It is interesting to find exact
analytic solutions for general $\lambda$ and  $\beta$. We also write
out the Friedmann equation for cosmological solutions in the $z=4$
Ho${\rm\check{r}}$ava-Lifshitz gravity in 3+1 dimensions for general
$\lambda$ and $\beta$, which shows that for $k=0$ there is no
contribution from $z=4$ terms.

The quantum inheritance principle is very important in the
construction of the Ho${\rm\check{r}}$ava-Lifshitz gravity and still
needs to be further understood in the framework of quantum gravity.
A relevant problem is the relationship between the new massive
gravity and the Ho${\rm\check{r}}$ava-Lifshitz gravity, which also
needs to be further studied. If the Ho${\rm\check{r}}$ava-Lifshitz
gravity is indeed a quantum gravity, the understanding of AdS/CFT
correspondence in the frame work of this theory would be very
important and it would be interesting to further analyze the
asymptotically AdS solutions in five dimensions we obtained in this
paper. In addition, it is very interesting to study  the
implications of this gravity theory in cosmology.

\section*{Acknowledgments}
We would like to thank  Hong L${\rm \ddot{u }}$ for useful
correspondence. This work was supported in part by the Chinese
Academy of Sciences with Grant No. KJCX3-SYW-N2 and the NSFC with
Grant No. 10821504 and No. 10525060.


\begin{thebibliography}{}

\bibitem{Horava:2008jf}
  P.~Horava,
  ``Quantum Criticality and Yang-Mills Gauge Theory,''
  arXiv:0811.2217 [hep-th].

\bibitem{Horava:2008ih}
  P.~Horava,
  ``Membranes at Quantum Criticality,''
  JHEP {\bf 0903}, 020 (2009)
  [arXiv:0812.4287 [hep-th]].

\bibitem{Horava:2009uw}
  P.~Horava,
  ``Quantum Gravity at a Lifshitz Point,''
  Phys.\ Rev.\  D {\bf 79}, 084008 (2009)
  [arXiv:0901.3775 [hep-th]].

\bibitem{Horava:2009if}
  P.~Horava,
  ``Spectral Dimension of the Universe in Quantum Gravity at a Lifshitz
  Point,''
  arXiv:0902.3657 [hep-th].


\bibitem{Takahashi:2009wc}
  T.~Takahashi and J.~Soda,
  ``Chiral Primordial Gravitational Waves from a Lifshitz Point,''
  arXiv:0904.0554 [hep-th].

\bibitem{Calcagni:2009ar}
  G.~Calcagni,
  ``Cosmology of the Lifshitz universe,''
  arXiv:0904.0829 [hep-th].

\bibitem{Kiritsis:2009sh}
  E.~Kiritsis and G.~Kofinas,
  ``Horava-Lifshitz Cosmology,''
  arXiv:0904.1334 [hep-th].

\bibitem{Kluson:2009sm}
  J.~Kluson,
  ``Branes at Quantum Criticality,''
  arXiv:0904.1343 [hep-th].

\bibitem{Lu:2009em}
  H.~Lu, J.~Mei and C.~N.~Pope,
  ``Solutions to Horava Gravity,''
  arXiv:0904.1595 [hep-th].

\bibitem{Mukohyama:2009gg}
  S.~Mukohyama,
  ``Scale-invariant cosmological perturbations from Horava-Lifshitz gravity
  without inflation,''
  arXiv:0904.2190 [hep-th].

\bibitem{Brandenberger:2009yt}
  R.~Brandenberger,
  ``Matter Bounce in Horava-Lifshitz Cosmology,''
  arXiv:0904.2835 [hep-th].

\bibitem{Nikolic:2009jg}
  H.~Nikolic,
  ``Horava-Lifshitz gravity, absolute time, and objective particles in curved
  space,''
  arXiv:0904.3412 [hep-th].



\bibitem{Nastase:2009nk}
  H.~Nastase,
  ``On IR solutions in Horava gravity theories,''
  arXiv:0904.3604 [hep-th].


\bibitem{Cai:2009pe}
  R.~G.~Cai, L.~M.~Cao and N.~Ohta,
  ``Topological Black Holes in Horava-Lifshitz Gravity,''
  arXiv:0904.3670 [hep-th].

\bibitem{Piao:2009ax}
  Y.~S.~Piao,
  ``Primordial Perturbation in Horava-Lifshitz Cosmology,''
  arXiv:0904.4117 [hep-th].

\bibitem{Gao:2009bx}
  X.~Gao,
  ``Cosmological Perturbations and Non-Gaussianities in Ho\v{r}ava-Lifshitz
  Gravity,''
  arXiv:0904.4187 [hep-th].

\bibitem{Ambjorn:2005db}
  J.~Ambjorn, J.~Jurkiewicz and R.~Loll,
  ``Spectral dimension of the universe,''
  Phys.\ Rev.\ Lett.\  {\bf 95}, 171301 (2005)
  [arXiv:hep-th/0505113].


\bibitem{Deser:1981wh}
  S.~Deser, R.~Jackiw and S.~Templeton,
  ``Topologically massive gauge theories,''
  Annals Phys.\  {\bf 140}, 372 (1982)
  [Erratum-ibid.\  {\bf 185}, 406.1988\ APNYA,281,409 (1988\ APNYA,281,409-449.2000)].

\bibitem{Deser:1982vy}
  S.~Deser, R.~Jackiw and S.~Templeton,
  ``Three-Dimensional Massive Gauge Theories,''
  Phys.\ Rev.\ Lett.\  {\bf 48}, 975 (1982).

\bibitem{Li:2008dq}
  W.~Li, W.~Song and A.~Strominger,
  ``Chiral Gravity in Three Dimensions,''
  JHEP {\bf 0804}, 082 (2008)
  [arXiv:0801.4566 [hep-th]].

\bibitem{Bergshoeff:2009hq}
  E.~A.~Bergshoeff, O.~Hohm and P.~K.~Townsend,
  ``Massive Gravity in Three Dimensions,''
  arXiv:0901.1766 [hep-th].

\bibitem{Nakasone:2009bn}
  M.~Nakasone and I.~Oda,
  ``On Unitarity of Massive Gravity in Three Dimensions,''
  arXiv:0902.3531 [hep-th].


\bibitem{Clement:2009gq}
  G.~Clement,
  ``Warped $AdS_3$ black holes in new massive gravity,''
  arXiv:0902.4634 [hep-th].
  W.~Kim and E.~J.~Son, ``Central Charges in 2d Reduced Cosmological Massive Gravity,''
  arXiv:0904.4538 [hep-th].




\bibitem{Liu:2009bk}
  Y.~Liu and Y.~Sun,
  ``Note on New Massive Gravity in $AdS_3$,''
  arXiv:0903.0536 [hep-th].


\bibitem{Nakasone:2009vt}
  M.~Nakasone and I.~Oda,
  ``Massive Gravity with Mass Term in Three Dimensions,''
  arXiv:0903.1459 [hep-th].


\bibitem{Liu:2009kc}
  Y.~Liu and Y.~W.~Sun,
  ``Consistent Boundary Conditions for New Massive Gravity in $AdS_3$,''
  arXiv:0903.2933 [hep-th].


\bibitem{Liu:2009ph}
  Y.~Liu and Y.~W.~Sun,
  ``On the Generalized Massive Gravity in $AdS_3$,''
  arXiv:0904.0403 [hep-th].


\bibitem{AyonBeato:2009yq}
  E.~Ayon-Beato, G.~Giribet and M.~Hassaine,
  ``Bending AdS Waves with New Massive Gravity,''
  arXiv:0904.0668 [hep-th].


\bibitem{Oda:2009ys}
  I.~Oda,
  ``Renormalizability of Massive Gravity in Three Dimensions,''
  arXiv:0904.2833 [hep-th].

\bibitem{CS} R.~G.~Cai and K.~S.~Soh,
  ``Topological black holes in the dimensionally continued gravity,''
  Phys.\ Rev.\  D {\bf 59}, 044013 (1999)
  [arXiv:gr-qc/9808067].

\bibitem{BTZ}M.~Banados, C.~Teitelboim and J.~Zanelli,
  ``Dimensionally continued black holes,''
  Phys.\ Rev.\  D {\bf 49}, 975 (1994)
  [arXiv:gr-qc/9307033].

\bibitem{Cai}R.~G.~Cai,
  ``Gauss-Bonnet black holes in AdS spaces,''
  Phys.\ Rev.\  D {\bf 65}, 084014 (2002)
  [arXiv:hep-th/0109133].


\end{thebibliography}
\end{document}